\documentclass[twocolumn]{aastex631}

\usepackage{color}

\received{}
\revised{}
\accepted{}

\submitjournal{ApJ}

\shorttitle{Supernova Precursors}
\shortauthors{Tsuna, Takei, Shigeyama}

\begin{document}

\title{Precursors of Supernovae from Mass Eruption: Prospects for Early Warning of Nearby Core-collapse Supernovae}

\correspondingauthor{Daichi Tsuna}
\email{tsuna@resceu.s.u-tokyo.ac.jp}

\author[0000-0002-6347-3089]{Daichi Tsuna}
 \affiliation{Research Center for the Early Universe, Graduate School of Science, University of Tokyo, Bunkyo-ku, Tokyo 113-0033, Japan}

\author[0000-0002-8215-5019]{Yuki Takei}
\affiliation{Research Center for the Early Universe, Graduate School of Science, University of Tokyo, Bunkyo-ku, Tokyo 113-0033, Japan}
\affiliation{Astrophysical Big Bang Laboratory, RIKEN, 2-1 Hirosawa, Wako, Saitama 351-0198, Japan}

\author[0000-0002-4060-5931]{Toshikazu Shigeyama}
\affiliation{Research Center for the Early Universe, Graduate School of Science, University of Tokyo, Bunkyo-ku, Tokyo 113-0033, Japan}
\affiliation{Department of Astronomy, School of Science, The University of Tokyo, 7-3-1 Hongo, Bunkyo-ku, Tokyo 113-0033, Japan}

\begin{abstract}
Recent observations of a large fraction of Type II supernovae show traces of dense circumstellar medium (CSM) very close to the progenitor star. If this CSM is created by eruptive mass loss several months before core-collapse, the eruption itself may be visible as a precursor, helpful as an early warning of a near-future supernova. Using radiation hydrodynamical simulations based on the open-source code CHIPS, we theoretically model the emission from mass eruption of a red supergiant star. We find that for a modest mass eruption the luminosity is typically on the order of $10^{39}\ {\rm erg\ s^{-1}}$, can last as long as hundreds of days until the star explodes, and is mainly bright in the infrared (from $-9$ to $-11$ mag around peak). We discuss observational strategies to find these signatures from Galactic and local Type II supernovae.
\end{abstract}

\keywords{supernovae: general --- stars: mass-loss --- circumstellar matter --- transient sources}

\section{Introduction}
The detections of photons and neutrinos \citep{Hirata87,Bionta87,Alexeyev88} from SN 1987A was a landmark for multi-messenger astronomy. With current detectors of electromagnetic and gravitational waves as well as neutrinos, future Galactic/local supernovae (SNe) would dramatically deepen our understanding of stellar core-collapse \citep[e.g.,][]{Andersson13,Adams13,Horiuchi18,Szczepanczyk21}.

Type II SNe, which are explosions of massive stars with a hydrogen-rich envelope, constitute a majority ($50$--$70$\%; \citealt{Li11,Smith11}) of these core-collapse SNe. Recent studies based on the early phase light curves \citep{Morozova17, Das17,Morozova18,Forster18} and spectra \citep{Yaron17,Boian20,Bruch21} of these SNe indicate a common presence of dense circumstellar medium (CSM) close to the star. The high density cannot be explained by standard wind mass-loss. 

The origins of the dense CSM have been theoretically investigated by various works. Many of the proposed mechanisms, such as binary interaction \citep{Chevalier12}, turbulent convection at the core \citep{Smith_Arnett14}, and wave heating \citep[e.g.,][]{Quataert12,Shiode14,Fuller17,Wu21}, involve energy injection in the stellar envelope that triggers eruptive mass loss \citep{Dessart10,Kuriyama20,Leung20,Matzner21,Ko22}\footnote{An alternative scenario to create overdense shell-like CSM by colliding-wind binaries was suggested by \cite{Kochanek19}, but such systems may be too rare ($<1\%$) to explain the entirety of Type II SNe \citep{Pejcha22}.}.

For Type IIn SNe that are believed to have extended massive CSM \citep{Smith14}, such pre-SN eruptions are indeed observed months to years before the final explosion \citep{Fraser13,Ofek14,Strotjohann21}. For Type II-P SNe that are more common, pre-SN eruptions are constrained by the monitoring of nearby red supergiants (RSGs). From the non-detection of strong variability from four Type II SN progenitors, \cite{Johnson18} found that eruptions after oxygen ignition in the core (a few years before core-collapse) are not typical, with probability $<37\%$. However the limiting cadence still allows eruptions within months before core-collapse. Indeed a precursor event was recently observed from $\approx 100$ days before explosion for a Type II-P SN 2020tlf \citep{Jacobson22}. Furthermore, from observations of Type II-P progenitors within 10 years of collapse, \cite{Davies22} claims that such brief outbursts are favorable as origin of the dense CSM.

Though the first signal of a nearby core-collapse SN is usually presumed to be neutrinos, this may not be the case for hydrogen-rich SNe if such pre-SN eruptions are ubiquitous. In this work we consider emission from pre-SN eruption(s), and estimate the feasibility of this as an {\it early warning} of a near-future SN event. From simulations of mass eruptions from an RSG progenitor, we find that the emission is bright in the infrared with luminosity of $\sim 10^6\ L_\odot$, potentially lasting for hundreds of days until the SN explosion.

This paper is constructed as follows. In Section \ref{sec:methods} we describe our formulations for calculating the mass eruption and the corresponding precursor emission. In Section \ref{sec:results} we present bolometric and multi-band light curves for various cases of energy injection. In Section \ref{sec:prospects} we compare our results to previous detections of precursor emission, and discuss strategies to find these events by optical/infrared surveys. We conclude in Section \ref{sec:conclusions}.

\section{Methods}
\label{sec:methods}

\begin{table*}
\centering
\begin{tabular}{c||cc|ccc}
Model name & Energy Injection & Parameters & Unbound mass [${\rm M}_\odot$] & $E_{\rm kin}$ [erg] & $E_{\rm rad}$ [erg]\\ \hline
single-fid & & $f_{\rm inj}=0.5$ & $0.35$& $2\times 10^{46}$ & $4\times 10^{46}$\\
single-small & once & $f_{\rm inj}=0.3$ &$0.015$& $2\times 10^{45}$ & $2\times 10^{46}$\\
single-large && $f_{\rm inj}=0.8$ &$1.2$& $7\times 10^{46}$ & $8\times 10^{46}$\\
\hline
double-fid &&  $f_{\rm inj,1}=0.3, f_{\rm inj,2}=0.5, \Delta t_{\rm inj}=100\ {\rm day}$ & 1.3&$4\times 10^{46}$ & $9\times 10^{46}$\\
double-long &twice& $f_{\rm inj,1}=0.3, f_{\rm inj,2}=0.5, \Delta t_{\rm inj}=200\ {\rm day}$&1.2&$4\times 10^{46}$ & $7\times 10^{46}$\\
double-large && $f_{\rm inj,1}=0.3, f_{\rm inj,2}=0.8, \Delta t_{\rm inj}=100\ {\rm day}$&3.6&$1.5\times 10^{47}$ & $1.4\times 10^{47}$
\end{tabular}
\caption{The models and parameters for mass eruption simulated in this work. Here $f_{\rm inj}$ is the injected thermal energy scaled by the initial total energy of the envelope, and $\Delta t_{\rm inj}$ is the interval between two injections. For the single injection models, we also show the mass of the unbound CSM, kinetic energy of the CSM, and the total radiated energy at the end of the simulation.}
\label{table:Parameters}
\end{table*}

\subsection{Model Setup}
As a representative progenitor of Type II SN, we adopt a RSG model of initial mass $15\ M_\odot$ and metallicity $0.014$, generated by the stellar evolution code MESA \citep{Paxton11,Paxton13,Paxton15,Paxton18,Paxton19} revision 12778 \footnote{In Appendix \ref{sec:other_progenitors} we consider eruptions of two other RSG progenitors, finding that the general characteristics of the precursor emission is similar.}. We use the progenitor model at core-collapse, that has a radius of $R_*=670\ R_\odot$, mass of $12.8M_\odot$, luminosity of $1\times 10^5L_\odot$, and effective temperature of $4000$ K. We inject thermal energy into the base of the progenitor's hydrogen-rich envelope, and simulate the resulting mass eruption.

We consider two cases for the energy injection, where the injection occurs once and twice. This results in one and two mass eruptions, respectively. The latter possibility may be not only theoretically expected due to distinct phases of nuclear burning with significant energy transported to the envelope via e.g. waves \citep{Shiode14,Wu21,Leung21}, but also from observations of some Type IIn SNe, such as SN 2009ip \citep{Mauerhan13,Pastorello13}. In fact, the main motivation for considering multiple mass eruptions is that the emission following the second eruption can be potentially much brighter than the first. Collision between the second ejecta and the slower part of the first ejecta may efficiently convert the kinetic energy of the second ejecta to radiation, analogous to Type IIn SNe \citep[e.g.,][]{Chugai91,Aretxaga99,vanmarle10,Chevalier11,Moriya13,Murase14,Tsuna19,Kuriyama21}.

The parameter sets we consider are summarized in Table \ref{table:Parameters}. The key parameter is $f_{\rm inj}$, the injected energy normalized by the initial binding energy of the hydrogen-rich envelope. The binding energy is $4.86\times 10^{47}\ {\rm erg}$ for this $15\ M_\odot$ RSG progenitor, and unbound material is ejected for energy injection of $\gtrsim 8\times 10^{46}\ {\rm erg}$, or $f_{\rm inj}\gtrsim 0.17$ \citep{Ko22}. We consider the range $0.3$--$0.8$, which results in an amount of CSM consistent with the range inferred from light curve modelling of Type II SNe \citep{Morozova18}. For the models with two eruptions, we additionally introduce a parameter $\Delta t_{\rm inj}$, the interval between two injections. For all cases, energy is injected impulsively in $10^3$ seconds with constant rate \footnote{We study cases of longer injection duration in Appendix \ref{sec:duration}, as the dynamics can be affected for duration comparable to the dynamical timescale of the envelope \citep{Ko22}. The consequence of a longer duration is effectively similar to a lower $f_{\rm inj}$.}.

\subsection{Mass Eruption Calculation}
 \begin{figure*}
   \centering
    \begin{tabular}{cc}
     \begin{minipage}[t]{0.5\hsize}
    \centering
    \includegraphics[width=\linewidth]{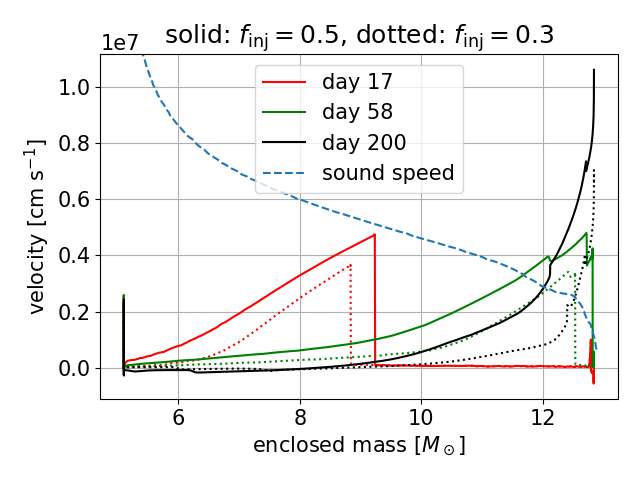}
    \end{minipage}
     \begin{minipage}[t]{0.5\hsize}
   \centering
    \includegraphics[width=\linewidth]{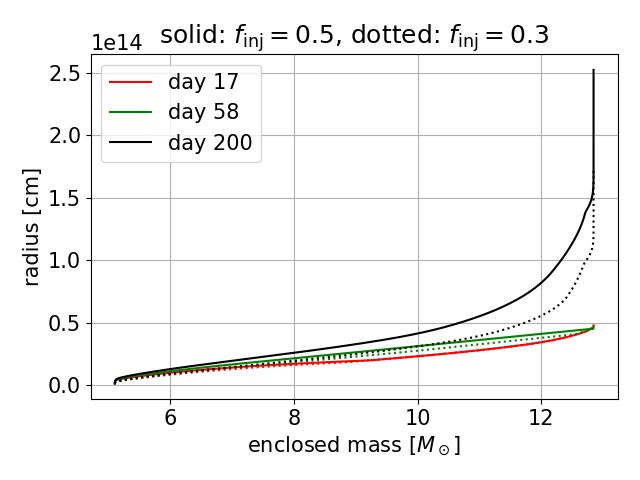} 
    \end{minipage} \\
    \begin{minipage}[t]{0.5\hsize}
    \centering
    \includegraphics[width=\linewidth]{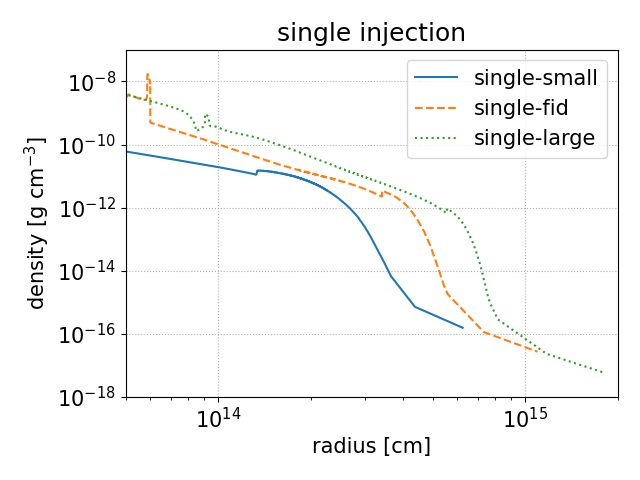}
    \end{minipage}
     \begin{minipage}[t]{0.5\hsize}
   \centering
    \includegraphics[width=\linewidth]{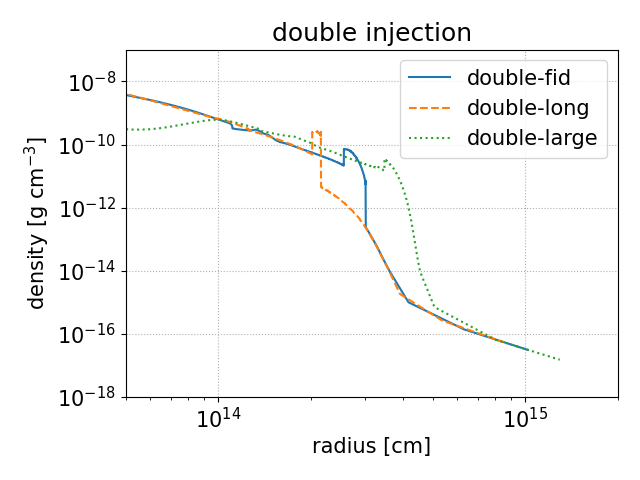} 
    \end{minipage}
    \end{tabular}
\caption{Summary of the mass eruption and resulting dense CSM. Top panels show the velocities and radii of the envelope as a function of enclosed mass. Solid lines are from the single-fid model, and dotted lines are from the single-small model. The dashed line is the sound-speed profile before energy injection. Bottom panels show the density profile at the end of the simulation, for the single and double injection models.}
 \label{fig:eruption}
 \end{figure*}

Our calculation of mass eruption is based on the one-dimensional radiation hydrodynamical simulation code from \cite{Kuriyama20}, that is public as part of the open-source code CHIPS \citep{Takei22}. The simulation follows the hydrodynamical evolution of the hydrogen-rich envelope after energy injection assuming local thermodynamic equilibrium. Radiative transfer is included by flux-limited diffusion \citep{Levermore81} with a grey opacity $\kappa(\rho,T)$, where $\rho$ and $T$ are respectively the gas density and temperature. For each model, we follow the evolution of the envelope for 600 days from energy injection.

We have made three important updates to the code. First, we have updated the analytical formula of $\kappa$ used in \cite{Kuriyama20} to a tabulated opacity covering a wide temperature range of $1600$ K$<T<10^7$ K.
 The analytical formula greatly simplifies the contribution from molecules and is generally overestimated at low temperatures of $T<5000$ K, with differences of at most two orders of magnitude. This regime is relevant in the outer regions of RSGs and can significantly impact the emission as well as the location and temperature of the photosphere. We have thus updated the formulation to obtain $\kappa$ by using two publicly available Rosseland-mean opacity tables, OPAL \citep{Iglesias96} and {\AE}SOPUS 1.0 \citep{Marigo09}\footnote{The {\AE}SOPUS table was recently updated to version 2.0 \citep{Marigo22}. For a wide range of gas density, only the opacities at temperatures below $2000$ K is affected, with differences within a factor of $2$. As this is lower than the temperature range important for our work, using the updated tables affects our light curves by only a tiny amount ($\sim 1\%$ near peak) with respect to version 1.0.}. These cover temperature ranges of $6000$--$10^7$ K and $1600$--$3\times 10^4$ K, respectively. We first generate the two opacity tables for the mass-averaged abundance in the envelope of $X=0.685, Y=0.301, Z=0.014$. Both of them tabulate $\kappa$ by $T$ and $R\equiv 10^{18}\rho/T^3$ in the range of $10^{-8}$--$10\ {\rm g\ cm^{-3}\ K^{-3}}$. We stitch the two tables at $10^4$ K, where the values of $\kappa$ generally agree within 10\% at all values of $R$. In rare cases that values of $\rho$ or $T$ outside the table are requested in the simulation, we use the edge values.

Secondly, we treat ionization more rigorously than \cite{Kuriyama20} by solving the Saha equations for the ionization degrees of hydrogen and helium ($x_{\rm H}$, $x_{\rm He}$, $x_{\rm He^+}$) \footnote{We neglect the ionization of metals. For our RSG model the metallicity is close to solar and the contribution of free electrons from metals is thus minor.}. The equation of state is replaced by the following two equations from the HELMHOLTZ equation of state that assumes completely ionized gas \citep{Timmes99,Timmes00}
\begin{eqnarray}
    p &=& \frac{\rho \mathcal{R}T}{\mu}+\frac{aT^4}{3} \label{eq:eos_p},\\
    e_{\rm int} &=& \frac{3}{2}\frac{\mathcal{R}T}{\mu}+\frac{aT^4}{\rho}+e_{\rm ion},
\end{eqnarray}
where $\mu$ is the mean molecular weight obtained from the Saha equations, $p$ is the pressure, $e_{\rm int}\ [{\rm erg\ g^{-1}}]$ is the specific internal energy, $\mathcal{R}$ is the gas constant, and $a$ is the radiation constant. $e_{\rm ion}(\leq 0)$ is the ionization energy defined as
\begin{eqnarray}
    e_{\rm ion}\equiv \frac{X(x_{\rm H}-1)\chi_{\rm H}}{m_{\rm H}} + \frac{Y[x_{\rm He}\chi_{\rm He}+(x_{\rm He^+}-1)\chi_{\rm He^+}]}{4m_{\rm H}},
\end{eqnarray}
where $\chi_{\rm H}\approx13.6{\rm eV}, \chi_{\rm He}\approx24.6{\rm eV},  \chi_{\rm He^+}\approx54.4{\rm eV}$ are the ionization energies of hydrogen and helium. The original code by \cite{Kuriyama20} treats the gas as completely ionized, i.e. $\mu\approx 0.62$ for solar composition and $e_{\rm ion}=0$. As discussed later, this update results in substantial enhancement on the late phase emission, that is essentially powered by recombination of hydrogen in the ejecta.

Thirdly, we attached an atmosphere to the outer edge of the progenitor. We adopt the Eddington grey relation\footnote{\url{https://docs.mesastar.org/en/latest/atm/t-tau.html}} $T^4(\tau)=0.75T_{\rm eff}^4(\tau+2/3)$ to approximate the temperature structure. Here the opacity in the atmosphere is assumed to be constant, with a value at the outer edge of the MESA model. We use the hydrostatic equilibrium, the equation of state (equation \ref{eq:eos_p}), and the continuity equation to respectively solve for the remaining quantities $p, \rho$ and $r$.
 
An example of mass eruption is shown in Figure \ref{fig:eruption}. Once energy is injected a shock forms and propagates in the star. The shock runs for a few months until it reaches the surface and breaks out. After shock breakout, the outermost material is accelerated to $100\ {\rm km\ s^{-1}}$ and becomes unbound from the RSG progenitor. Some part of the ejected envelope is still bound to the envelope, and eventually falls back onto the star. The mass of the unbound CSM is sensitive to $f_{\rm inj}$, as shown in Table \ref{table:Parameters}.

The density profile of the erupted envelope is shown in the bottom panels of Figure \ref{fig:eruption}. A notable point is that while the single injection models have a smooth double power-law profile, the double injection models display a high-density region at a few $\times 10^{14}$ cm. This corresponds to the shocked region created by the interaction between the first and second ejecta  \citep{Chevalier82}, analogous to Type IIn SNe.

\begin{figure}
    \centering
    \includegraphics[width=\linewidth]{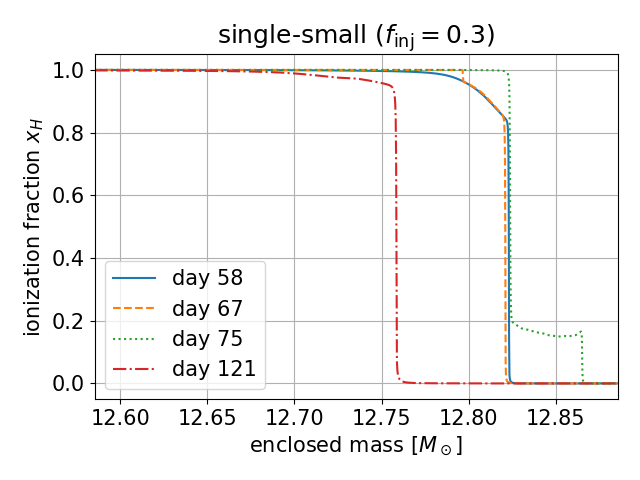}
 \caption{Evolution of the hydrogen ionization fraction at the outermost part of the envelope, for the single-small model.}
\label{fig:hyd_fraction}
\end{figure}
In Figure \ref{fig:hyd_fraction} we show the time evolution of $x_H$ at the outermost part of the envelope, focusing on around and after shock breakout for the case of $f_{\rm inj}=0.3$. The shock heats and ionizes the envelope (day 58 to 67) until it breaks out. The escaping radiation after breakout also partially ionizes the upstream (day 75). The ejected envelope later cools down, and recombines from the outside.

\subsection{Light Curve Calculation}
\label{sec:lightcurve}
We use the simulation results to calculate both bolometric light curves and the multi-band light curves. For the bolometric light curve, we simply extract the luminosity flowing through the outermost cell. Although the outermost cell moves at a finite velocity, it moves much slower than the speed of light and light-travel time effects can be neglected. The luminosity is extracted at intervals of $0.1$ day, which is roughly equivalent to the shock crossing time in the outermost cells. 

To obtain the multi-band light curves, we first obtain the location of the photosphere from the equation 
\begin{equation}
T(r)=T_{\rm eff}(r)\equiv \left(\frac{L(r)}{4\pi r^2\sigma_{\rm SB}}\right)^{1/4},
\label{eq:T_eff}
\end{equation} 
where $\sigma_{\rm SB}$ is the Stefan-Boltzmann constant, and profiles $T(r)$ and $L(r)$ are extracted from our simulation. We assume that the emission is thermal, with temperature and luminosity at the photosphere. This assumption is reasonable because the temperature outside the photosphere is $\lesssim 5000$ K, where absorption is a more dominant source of opacity than Thomson scattering. Thus we can approximate the effective temperature in equation (\ref{eq:T_eff}) as the color temperature. Using the spectral synthesis code CLOUDY \citep{Ferland17}, we have checked that the gas outside the photosphere does not affect the spectra, except at very long wavelengths of $\lambda \gtrsim 5\mu$m. 

We have used the public filter transmission functions by the Roman Space Telescope\footnote{\url{https://roman.gsfc.nasa.gov/science/Roman_Reference_Information.html}} available from optical to near-infrared bands. The four bands we selected, F062, F106, F129, and F158, have central wavelengths similar to the R, Y, J, and H bands respectively.
Finally, we focus only after the shock breaks out from the star, when the fluid can be assumed as steady state. 
\section{Results}
\label{sec:results}

\subsection{Single Eruption Case}
\begin{figure*}
   \centering
    \begin{tabular}{cc}
     \begin{minipage}[t]{0.5\hsize}
    \centering
    \includegraphics[width=\linewidth]{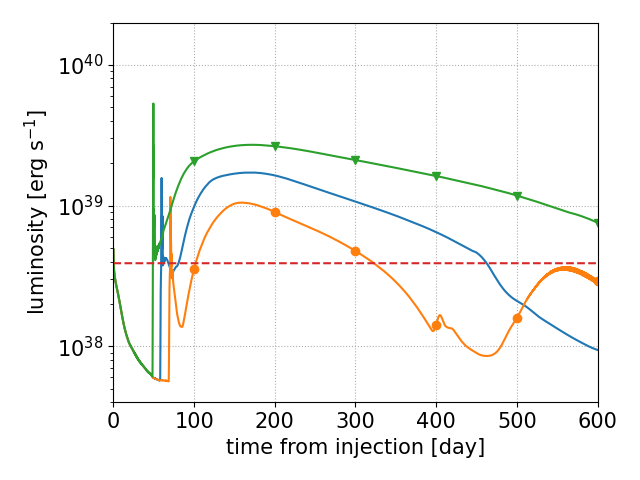}
    \end{minipage}
     \begin{minipage}[t]{0.5\hsize}
   \centering
    \includegraphics[width=\linewidth]{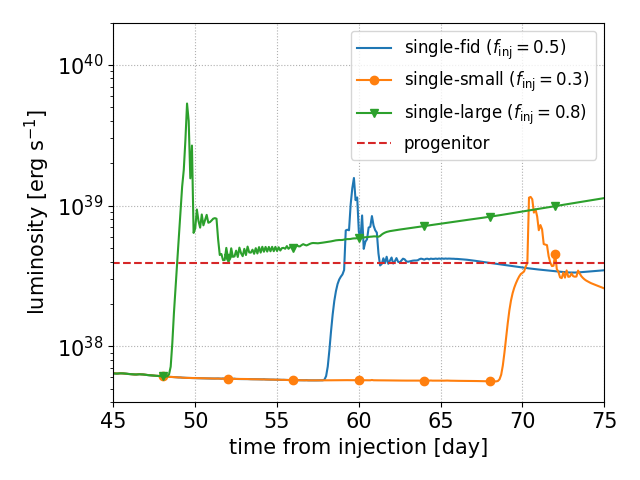}
    \end{minipage}
    \end{tabular}
 \caption{Bolometric light curves for models with single energy injection. The horizontal dashed line shows the luminosity of the original RSG progenitor ($\approx 10^5L_\odot$).}
\label{fig:bolom}
\end{figure*}
Figure \ref{fig:bolom} shows the bolometric light curves in the single injection case for different values of $f_{\rm inj}$. Generally for large values of $f_{\rm inj}$ the overall emission is brighter. For all values of $f_{\rm inj}$ the light curve has two distinct peaks, the shock breakout emission of order days, and a subsequent cooling of the ejecta lasting for hundreds of days. For the $f_{\rm inj}=0.3$ model, we also find a peak at late phase that comes from fallback of the bound part of the envelope.

We note that the light curves are under the assumption that the SN has not yet occurred for the entire 600 days. However, monitoring of pre-SN RSGs \citep{Johnson18} find instead a quiescent progenitor up to $\approx 100$--$200$ days before the SN. Although we present the entire light curve for completeness, we would thus likely not see the entire light curve, but instead only the first $\lesssim 100$ days after shock breakout.

The initial decline, also seen in model light curves of \cite{Kuriyama20}, is a numerical artifact due to the simulation omitting convective heat transport. Without energy being convected from the inner region, the outermost layer radiatively cools and its luminosity and temperature thus drop. The resulting loss of internal energy before breakout ($\sim 10^{45}\ {\rm erg}$) is unlikely to affect the post-breakout emission, but the modified temperature distribution may affect the emission at breakout.

We pick up three characteristic aspects of a single eruption (the shock breakout, the role of recombination of hydrogen, and the fall back), and discuss them quantitatively in the following.
\subsubsection{Shock breakout}
A shock that propagates in the hydrogen-rich envelope heats the envelope and produces copious photons at the downstream. Initially the downstream photons are trapped there, but near the stellar surface photons can diffusively escape. This shock breakout \citep[e.g.,][]{Falk78,Matzner99,Waxman17} occurs at a radius that satisfies $\tau(r_{\rm bo})\approx c/v_{\rm sh}$, where $v_{\rm sh}$ is the shock velocity, $\tau$ is the optical depth from the surface to $r_{\rm bo}$, and $c$ is the speed of light. Because the optical depth quickly rises with depth from the surface, the width $\Delta R=R_*-r_{\rm bo}$ of the layer that radiation can efficiently escape is typically only a few \% of $R_*$ \citep{Ko22}. 

The timescale of the breakout emission is the diffusion time of photons in this breakout layer
\begin{eqnarray}
    t_{\rm bo}\approx \frac{\Delta R}{v_{\rm sh}} \sim 2\ {\rm day} \left(\frac{\Delta R}{10^{12}\ {\rm cm}}\right)\left(\frac{v_{\rm sh}}{60\ {\rm km\ s^{-1}}}\right)^{-1}.
    \label{eq:t_bo}
\end{eqnarray}
The total dissipated energy in this timescale can be crudely estimated as
\begin{eqnarray}
    E_{\rm bo}&\approx& \int^{R_*}_{r_{\rm bo}} 4\pi r^2 \left(\frac{1}{2}\rho(r) v_{\rm sh}^2\right)(v_{\rm sh}dt) \nonumber \\
    &\sim& 2\pi R_*^2v_{\rm sh}^2\left(\frac{c}{v_{\rm sh}\kappa}\right) \nonumber \\
    &\sim& 2\times 10^{44}\ {\rm erg} \nonumber \\
    && \times \left(\frac{\kappa}{10\ {\rm cm^2\ g^{-1}}}\right)^{-1}\left(\frac{R_*}{670R_\odot}\right)^2\left(\frac{v_{\rm sh}}{60\ {\rm km\ s^{-1}}}\right),
    \label{eq:E_bo}
\end{eqnarray}
where $v_{\rm sh}$ refers to the velocity at breakout, and we adopted $\kappa\approx 10\ {\rm cm^2\ g^{-1}}$ at the breakout radius (see also \citealt{Ko22}). These estimates for the breakout radiation can be subject to several uncertainties. First, this estimation neglects the evolution of the diffusion velocity of photons after the breakout. 
Second, for breakouts of a very slow shock like our case, the dissipated energy may not be dominated by radiation. The upstream density at breakout typically found from the simulations is $\rho_{\rm up}\approx 10^{-9}\ {\rm g\ cm^{-3}}$. The downstream temperature assuming adiabatic index of $5/3$ is
\begin{eqnarray}
    T_{\rm down}&=&\frac{3\mu m_p}{16k_B}v_{\rm sh}^2 \nonumber \\
    &\sim& 5\times 10^4\ {\rm K}\left(\frac{\mu}{0.63}\right)\left(\frac{v_{\rm sh}}{60\ {\rm km \ s^{-1}}}\right)^2.
    \label{eq:T_down}
\end{eqnarray}
The ratio of energy densities of radiation and gas is then
\begin{eqnarray}
    &&\frac{aT_{\rm down}^4}{1.5\cdot 4\rho_{\rm up}k_BT_{\rm down}/\mu m_p}\nonumber \\
    &\sim& 1.3\left(\frac{\rho_{\rm up}}{10^{-9}\ {\rm g\ cm^{-3}}}\right)^{-1} \left(\frac{\mu}{0.63}\right)^4\left(\frac{v_{\rm sh}}{60\ {\rm km \ s^{-1}}}\right)^6.
\end{eqnarray}
Thus we may assume that a significant fraction of $E_{\rm bo}$ is converted to radiation by the shock. If this ratio was much smaller than 1, the temperature of thermal photons would be reduced from that assuming radiation-dominated gas. This is in contrast to shock breakouts in typical SNe of $v_{\rm sh}\sim 10^4\ {\rm km\ s^{-1}}$, where the downstream has an order of magnitude higher temperature and is radiation dominated. As gas and radiation are coupled at these densities, the internal energy initially carried by gas is eventually released as radiation within a diffusion time ($\sim t_{\rm bo}$), and thus still contribute to the breakout emission. These processes would be important when modelling the spectra, which would likely peak in the ultraviolet.

To summarize, for typical parameters of the shock the timescale and radiated energy of the breakout pulse are a few days and $\sim 10^{44}$ ergs, respectively. These rough estimates reasonably agree with the simulated light curves in the right panel of Figure \ref{fig:bolom}. 

After the first peak of the breakout, the light curve flattens for a few days before it drops. This phase appears when the photosphere enters the denser layer after the outermost layer that was heated upon breakout cools down. At this point a recombination front develops and the light curve transitions to a recombination-powered phase, which is explained in the next section.

\subsubsection{Envelope Recombination}
\begin{figure*}
    \centering
    \includegraphics[width=\linewidth]{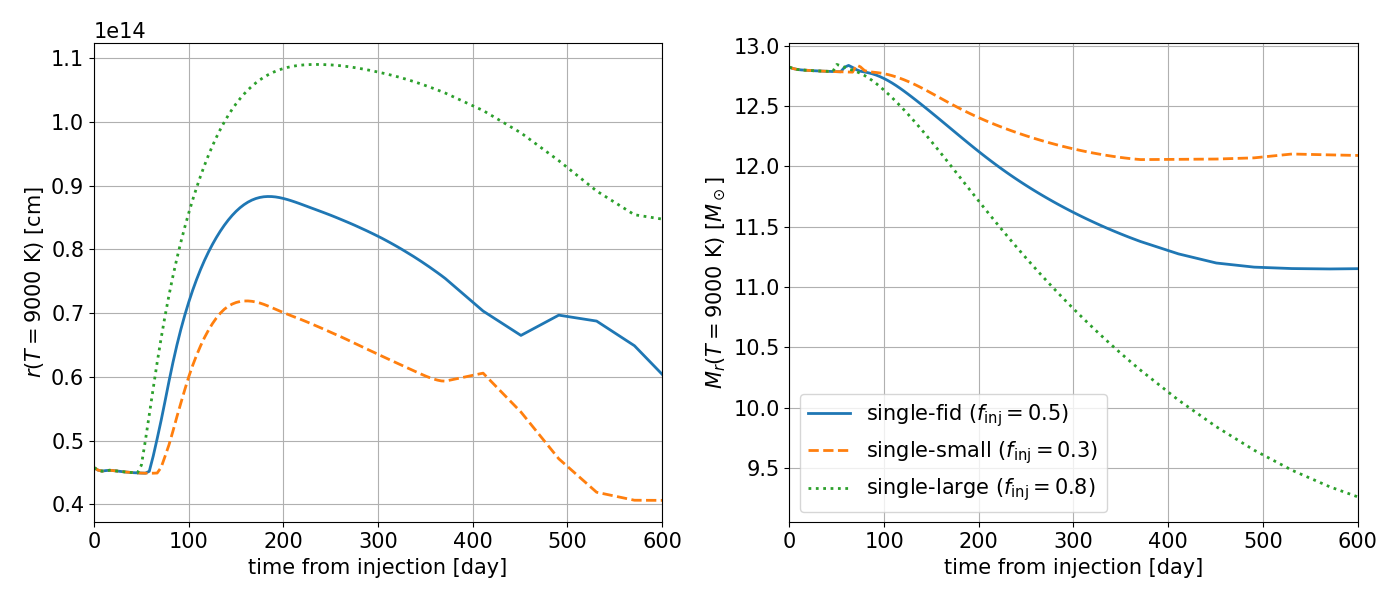}
 \caption{Radius and mass coordinate of the location where $T=9000$ K, approximately the temperature that hydrogen starts to recombine for a density of $10^{-10}\ {\rm g\ cm^{-3}}$.}
\label{fig:recomb}
\end{figure*}
The second peak, close to a plateau, is powered by cooling of the expelled envelope. This peak was not seen in the original work by \cite{Kuriyama20}, and we can presume that this emission is powered by the recombination of hydrogen neglected in their work. For $f_{\rm inj}=0.3$--$0.8$ the peak luminosity is $\approx 1$--$3\times 10^{39}\ {\rm erg\ s^{-1}}$. This is as bright as the breakout pulse, but much longer and hence easier to detect.

Figure \ref{fig:recomb} shows the location where $T=9000$ K, at which about half of hydrogen atoms recombine for a density of $\approx 10^{-10}\ {\rm g\ cm^{-3}}$. This location is equivalent to the recombination front that exists in Type II SNe \citep[e.g.,][]{Grassberg71,Grassberg76,Kasen09,Faran19}, except that the recombination temperature is higher due to the much higher gas density. Following shock breakout the recombination front first expands to $\sim 10^{14}$ cm, but starts to recede from day $\approx 200$.

Recombination of hydrogen at the front supplies an internal energy of 
\begin{eqnarray}
    \frac{13.6\ {\rm eV}}{m_p}XM_{\rm ej}\sim 2\times 10^{46}\ {\rm erg} \left(\frac{X}{0.7}\right)\left(\frac{M_{\rm ej}}{1\ M_\odot}\right),
    \label{eq:recomb_power}
\end{eqnarray}
where $M_{\rm ej}$ is the ejected mass. The right plot of Figure \ref{fig:bolom} shows the recession of the recombination front in mass coordinate. For $f_{\rm inj}=0.3$--$0.8$ the slopes of these curves range as $(0.5$--$1)M_\odot$/100 days, and from equation (\ref{eq:recomb_power}) this can power a luminosity of $1$--$2\times 10^{39}\ {\rm erg\ s^{-1}}$. These values are overall consistent with the peaks of the simulated light curves. The light curve of $f_{\rm inj}=0.8$ model, with a large explosion energy (Table \ref{table:Parameters}), probably has additional minor contribution from the internal energy of the envelope deposited by the shock. 
A much larger energy injection will enhance this contribution, but with a possible caveat that the entire hydrogen envelope may be expelled.

For the models of $f_{\rm inj}=0.3$ and $0.5$, the recombination front settles (in Lagrangian sense) at 300 and 450 days respectively. This is presumably when all the ejecta has recombined, and is consistent with the epoch where the light curve starts to drop steeply. 

In summary, hydrogen recombination can power a plateau emission of timescale $50$--$200$ days and luminosity of up to a few $\times 10^{39}\ {\rm erg\ s^{-1}}$. This emission, though much dimmer than the typical plateau observed in Type II SNe, is likely much easier to detect than the breakout pulse. 

\subsubsection{Envelope Fallback}
For $f_{\rm inj}=0.3$, a second dimmer bump that peaks at $\approx 550$ days is seen. This is due to the significant fallback in this model, as verified below.

The expanding envelope after mass eruption is gravitationally pulled by the central star. In a few dynamical timescales after mass eruption when internal energy becomes negligible, the acceleration is only by gravity,
\begin{eqnarray}
    \frac{dv}{dt}=-\frac{GM_r}{r^2}\sim -0.13\ {\rm cm\ s^{-2}}\left(\frac{M_r}{10\ {\rm M}_\odot}\right)\left(\frac{r}{10^{14}\ {\rm cm}}\right)^{-2}.
\end{eqnarray}
For the case of $f_{\rm inj}=0.3$, the slower part of the ejecta with velocity of $\sim 10\ {\rm km\ s^{-1}}$ (see Figure \ref{fig:eruption}) thus stalls an order of $100$ days after eruption. This stalling occurs much later for larger values of $f_{\rm inj}$, and thus does not appear in the light curve. After the envelope stalls, it falls back to the central star at a timescale
\begin{eqnarray}
    t_{\rm fb} \approx \sqrt{\frac{r^3}{GM_r}}\sim 300\ {\rm day}\left(\frac{r}{10^{14}\ {\rm cm}}\right)^{3/2}\left(\frac{M_r}{10\ {\rm M}_\odot}\right)^{-1/2}.
\end{eqnarray}
The fallback matter has a velocity of $v_{\rm fb}\approx-\sqrt{2GM_r/R_*}$. The density of the fallback matter at $r=R_*$ is $\rho_{\rm fb}\approx 1\times 10^{-10}\ {\rm g\ cm^{-3}}$ around the light curve peak\footnote{A similar value can also be obtained from the analytical modelling of \cite{Tsuna21} (see their Appendix A)}. The collision between the fallback matter and the star creates a shock, that dissipates the fallback matter's kinetic energy at a luminosity $2\pi R_*^2\rho_{\rm fb}(-v_{\rm fb}^3)\approx 6\times 10^{38}\ {\rm erg\ s^{-1}}$, which roughly agrees with the light curve. Afterwards this luminosity declines as the fallback rate drops as a function of time \citep{Tsuna21}.
\subsection{Double Eruption Case}
\label{sec:results_double}

 \begin{figure}
 \centering
 \includegraphics[width=\linewidth]{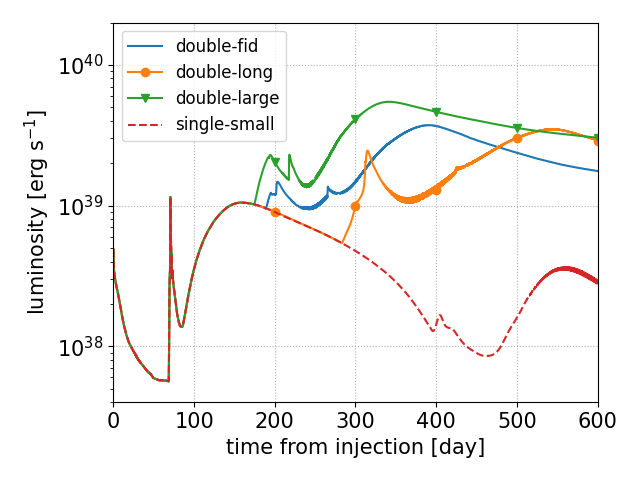}
\caption{Bolometric light curves for models with two energy injections. The dashed line shows the $f_{\rm inj}=0.3$ model in Figure \ref{fig:bolom}.}
 \label{fig:dbl_inj}
 \end{figure}

In Figure \ref{fig:dbl_inj} we show the bolometric light curves for two injections, overlapped with the single injection case of $f_{\rm inj}=0.3$. The second breakout occurs 50-100 days after the energy injection, followed by a rebrightening in the light curve.

In the second eruption, the shock sweeps through an inflated envelope with large radius and reduced density. Thus $\Delta R$ and $r_{\rm bo}$ are larger than the first shock breakout, and from equations (\ref{eq:t_bo}) and (\ref{eq:E_bo}) the emission has a longer duration and larger luminosity. This is more pronounced for larger $\Delta t_{\rm inj}$ which enables the first ejecta to expand farther. A larger $\Delta t_{\rm inj}$ also reduces the optical depth of the first ejecta, making photons easier to escape before suffering from adiabatic losses. These effects lead to the second breakout being much easier to observe than the first breakout.

The subsequent peak lasting for hundreds of days is again due to the cooling emission. For the double-fid and double-long models, the peak luminosity is both $4\times 10^{39}\ {\rm erg\ s^{-1}}$, slightly brighter than the single-large model that injected the same amount of total energy. For the double-large case with the largest energy injection, the peak luminosity is $\approx 6\times 10^{39}\ {\rm erg\ s^{-1}}$.

\subsection{Multi-band Light Curves}

 \begin{figure*}
   \centering
    \begin{tabular}{ccc}
     \begin{minipage}[t]{0.5\hsize}
    \centering
    \includegraphics[width=\linewidth]{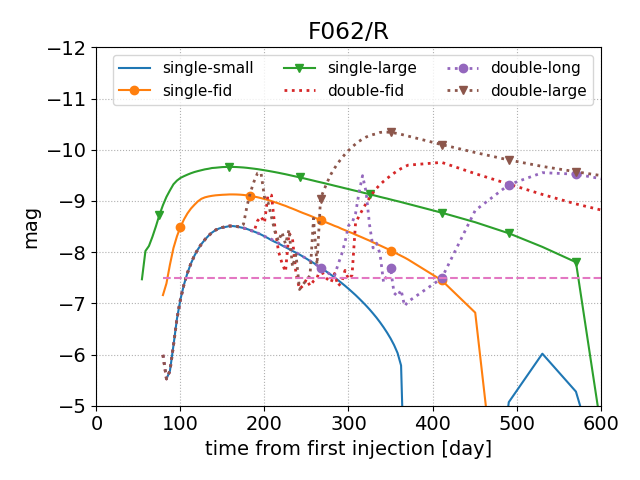}
    \end{minipage}
     \begin{minipage}[t]{0.5\hsize}
   \centering
    \includegraphics[width=\linewidth]{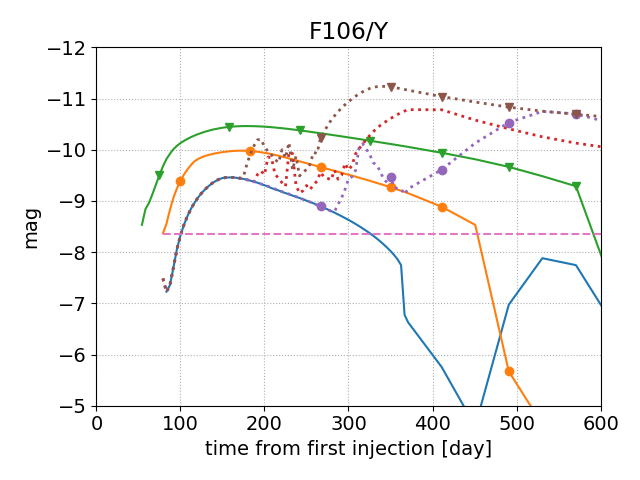}
    \end{minipage}
    \\
    \begin{minipage}[t]{0.5\hsize}
   \centering
    \includegraphics[width=\linewidth]{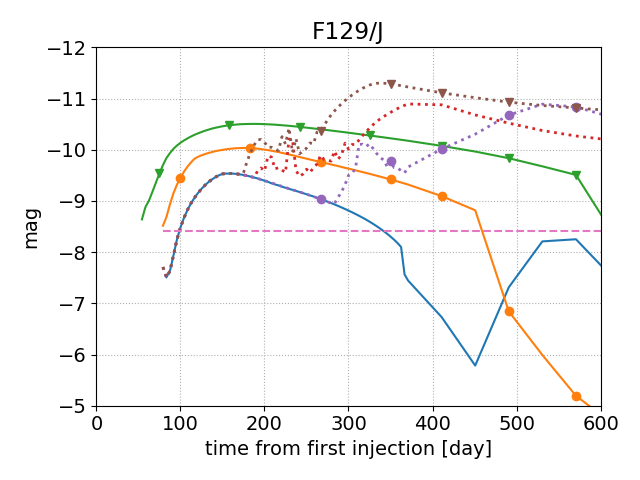}
    \end{minipage}
    \begin{minipage}[t]{0.5\hsize}
   \centering
    \includegraphics[width=\linewidth]{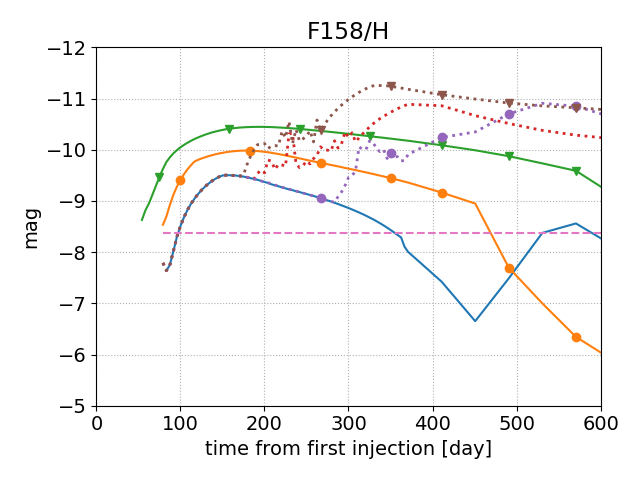}
    \end{minipage}
    \end{tabular}
\caption{Light curves in four different bands for the six model cases. The horizontal dashed line shows the magnitude of the original RSG progenitor. The four bands we selected, F062, F106, F129, and F158, have similar central wavelengths with the R, Y, J, and H bands respectively (see Section \ref{sec:lightcurve}).}
 \label{fig:multiband}
 \end{figure*}
 
Figure \ref{fig:multiband} shows the absolute magnitudes in four bands covering from optical to near-infrared. The single eruption models display a plateau of $\approx -9$ mag in the R-band and $\approx -10$ mag in the infrared bands. The emission at peak is 1-2 mag brighter than the original progenitor, shown as horizontal dashed lines.

The double eruption models display two peaks that arise from the second eruption. The second peak is typically brighter, with $\approx -11$ mag in the infrared. Despite the peak bolometric luminosity is similar to the single-large model, the peak magnitude in the infrared bands is brighter by up to $\approx 0.5$ mag. This is due to the larger radius of the photosphere for the double eruption models.

\section{Discussion}
\label{sec:prospects}
 
\subsection{Comparison with Observations}
\cite{Strotjohann21} have done the most extensive observational study of precursors, using SNe detected by the Zwicky Transient Facility. They have found months-long precursor events brighter than -13 mag in $\approx 25\%$ of Type IIn SNe. While this is much brighter than our predictions, the sensitivity of their observations leave the emission dimmer than $\approx -12$ mag unconstrained.

In fact, we speculate that these precursor events are unrelated to the eruptions of RSGs like considered in this work. First, the identified progenitors of Type IIn SNe with precursors, such as SN 2009ip and 2015bh, are found to be luminous blue variables rather than RSGs \citep{Smith10,Mauerhan13,Elias-rosa16,Boian18}. Second, the total radiated energy of these precursors are $10^{47}$--$10^{49}$ erg, which require at least this much energy deposited into the envelope. Since RSG envelopes have a low gravitational binding energy of order $10^{47}\ {\rm erg}$, this deposition results in most (if not all) of the envelope being ejected. While this may not be in conflict with Type IIn SNe, the subsequent SNe will be much brighter than typical Type II SNe due to CSM interaction \citep[e.g.,][]{Ouchi19}.

From the above arguments we conclude that eruptions of RSGs are not responsible for the detected precursors, at least for the brightest ones. Instead, eruptive mass loss from blue supergiant progenitors may be able to explain the energetics of the observed precursors, without having to expel the entire hydrogen-rich envelope. We plan to explore such possibility in future work. 

A mechanism not considered in this work is emission from a wind-like mass loss, where energy is continuously supplied to the envelope instead of instantaneously \citep{Quataert16}. This is claimed to explain a fraction of the precursors, such as the Type II-P SN 2020tlf (\citealt{Matsumoto22}; see also \citealp{Chugai22}). Highly asymmetric energy injection by e.g. jets (\citealt{Soker22} and references therein) may be another possibility to both realize the bright luminosity and retain most of the hydrogen-rich envelope. Our code assumes spherical symmetry, and studying these would require multi-dimensional simulations with radiation transfer.

A survey of nearby failed SNe observed an outburst of $\sim 10^6\ L_\odot$ from a RSG, followed by an apparent vanish of the progenitor in optical \citep{Gerke15,Adams17a}. In a failed SN, reduction of gravity by neutrino emission from the core can give an outward push to the envelope. This excess energy may expel the envelope \citep{Nadyozhin80,Lovegrove13,Fernandez18,Tsuna20,Ivanov21}, with effectively the same physics as considered here. While the observed luminosity roughly agrees with theoretical predictions of failed SNe of RSGs \citep{Lovegrove13,Fernandez18}, recent work based on a more sophisticated treatment of the inner core physics finds a dimmer value ($\lesssim 2\times 10^5L_\odot$; \citealt{Ivanov21}). This may imply the existence of a precursor event considered here, that can enhance the luminosity of the mass ejection.

\subsection{Prospects for Future Detection}
 \begin{figure}
 \centering
 \includegraphics[width=\linewidth]{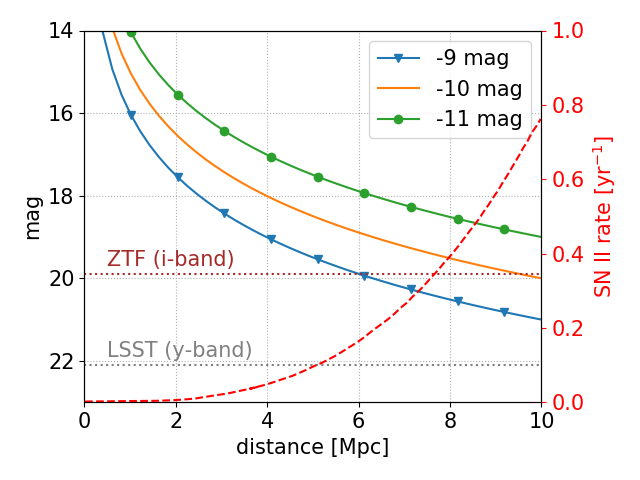}
\caption{The observed magnitude (solid lines) and all-sky Type II SN rate (dashed line) as a function of distance. We use the SN II rate from \cite{Kheirandish22}, which used the core-collapse SN rate indirectly inferred from nearby galaxies and assumed a SN II fraction of $60\%$. We note that the event rate of recent nearby core-collapse SNe is a few times larger than that obtained from indirect methods \citep{Nakamura16}. Horizontal dotted lines are the single-visit depths of ZTF \citep{Bellm19} and LSST \citep{Ivezic19} in the infrared bands.}
 \label{fig:prec_mag}
 \end{figure}
 
Given the sensitivity limit of current observations, we predict that there is a hidden population of dim precursors of Type II SNe that arise from eruptions of RSGs. Light curve modelling of Type II SNe constrain the extent of the dense CSM to be typically $\approx 2000\ R_\odot$ \citep{Morozova18}, which implies a timescale of $\lesssim 1$ year from eruption to core-collapse. This is long enough to assume that the light curve of the precursor can reach around its peak before the progenitor collapses.

In this section we consider observational strategies to find these events, focusing on local (within $10$ Mpc) and Galactic Type II SNe. The former distance range is motivated by the detectability by future Megaton-class neutrino facilities \citep{Ando05,Kistler11,Boser15}, and we can expect $\sim 1$ Type II SN per year within this volume \citep[e.g.,][]{Horiuchi13,Nakamura16}. Finding these precursors would enable detection of the (very brief) shock breakout of the following SN. This would help narrow down the temporal search window for neutrino and gravitational-waves, thus enhancing the sensitivity of these detectors.

For the local sources, a straightforward strategy is to monitor the nearby galaxies within 10 Mpc, preferentially in the near-infrared. Within 10 Mpc there are $\approx 40$ Milky Way-like galaxies that dominate the core-collapse SN rate. Given that outbursts occur order 100 days before the SN, monitoring these with a cadence of tens of days would realize early warning of the SN. 

In Figure \ref{fig:prec_mag} we plot the observed magnitude for precursors of $-11$ to $-9$ mag, along with the SN II rate. The required sensitivity is $\approx 20$--$21$ mag if one hopes to detect all of the precursors within 10 Mpc. This sensitivity is in fact expected to be achieved by the Rubin Observatory, with single-visit depths of $23.3$ and $22.1$ mag in the $z$ and $y$ bands respectively with cadence of 3 days \citep{Ivezic19}.

For Galactic SN precursors, an infrared search is likely required due to the severe extinction by foreground dust. A precursor at $\approx 10$ kpc is expected to be visible by naked eye ($\approx 6$ mag) without dust, but for a typical SN in the disk the emission will suffer extinction of up to 20--30 mag in the optical \citep{Adams13}. A near-infrared survey in e.g. the J-band will reduce the extinction by a factor of $3.5$ \citep{Cardelli89}, making searches more feasible. \cite{Adams13} finds that most (92\%) of the progenitor RSGs would have already been catalogued by 2MASS \citep{Skrutskie06}, so confusion with other sources can thus be mitigated by inspection of archival images.

\section{Conclusions}
\label{sec:conclusions}
Using the open-source code CHIPS, we have simulated the optical and infrared light curves of eruptions from a $15\ M_\odot$ RSG star. The light curves are powered by a brief shock breakout pulse of days, and a much longer cooling emission lasting for hundreds of days. For eruptions of lower energy, they are followed by a dim peak powered by fallback of the bound envelope.

The cooling emission can range from $-9$ to $-11$ mag in the near-infrared, and is brighter than the original RSG progenitor by several mag. We expect that precursors of local ($<10$ Mpc) and Galactic SNe can be found with surveys that have cadence of tens of days or shorter, preferably in the near infrared. We expect that these mass eruption events can serve as early warning of a near-future nearby SN, which will be important for multi-messenger studies of core-collapse SNe.

Monitoring of such eruptive events would also give us important clues for understanding the final moments of a massive star's life. The observed infrared/optical signals will give us clues to constrain the proposed mechanisms for extreme mass loss. On the other hand, quantitative predictions for the energy injections are limited at present (but see e.g. \citealt{Wu21}), and future work shall aim to determine the energy budget for progenitors of various masses.

\begin{acknowledgements}
We thank Christopher Kochanek, Tatsuya Matsumoto, Noam Soker, and the anonymous referee for constructive comments on this manuscript. This work is supported by JSPS KAKENHI Grant Numbers JP19J21578, JP20H05639, and JP21J13957, MEXT, Japan.
\end{acknowledgements}

\begin{appendix}

\section{Dependence on Red Supergiant Progenitor Models}
\label{sec:other_progenitors}

In this section we consider mass eruption from red supergiant progenitors with different zero-age main sequence (ZAMS) mass from the 15 $M_\odot$ model. Here we additionally construct two progenitors of ZAMS masses of $11\ M_\odot$ and $24\ M_\odot$, which likely lie at the extremes in the mass range of RSGs. These masses are within the mass ranges where the energy injection due to wave-heating can be relatively stronger \citep{Wu21}.

The progenitors were constructed by the test suite ``example\_make\_pre\_ccsn" available in MESA revision 12778, similar to the $15\ M_\odot$ model. The properties for the two new progenitors, along with our fiducial 15 $M_\odot$ progenitor, are shown in Table \ref{table:ProgenitorParameters}. For the $11\ M_\odot$ model, the evolution halts after core oxygen burning, due to numerical problems arising from a spike in the density profile just interior to the helium core. We estimate the time to core-collapse by dividing the total binding energy in the CO core by the neutrino luminosity, to be about 1 year. This is much shorter than the envelope's Kelvin-Helmholtz timescale $GM_*^2/R_*L_* \approx 100$ years, where $M_*$ and $L_*$ are respectively the mass and surface luminosity of the star. Thus we can approximate that the envelope is not greatly altered until core-collapse, and use this model in the eruption simulations. On the other hand the $24\ M_\odot$ model is successfully evolved up to core-collapse, and as in the $15\ M_\odot$ model we use the progenitor model at core-collapse in our simulations.

\begin{table*}
\centering
\begin{tabular}{c||cccc}
ZAMS mass [$M_\odot$] & Radius [$R_\odot$] & Effective temperature [K] & H-envelope mass [$M_\odot$] & Helium-core mass [$M_\odot$]\\ \hline
15 & 670 & 4000 & 7.9 & 4.9\\
11 & 460 & 4100 & 7.0 & 3.2\\
24 & 830 & 4400 & 8.3 & 8.4
\end{tabular}
\caption{The RSG progenitor models that we simulate in this Appendix. They are all generated by MESA revision 12778, with a metallicity of $0.014$ at zero-age main sequence (ZAMS).}
\label{table:ProgenitorParameters}
\end{table*}

\begin{figure*}
   \centering
    \begin{tabular}{cc}
     \begin{minipage}[t]{0.5\hsize}
    \centering
    \includegraphics[width=\linewidth]{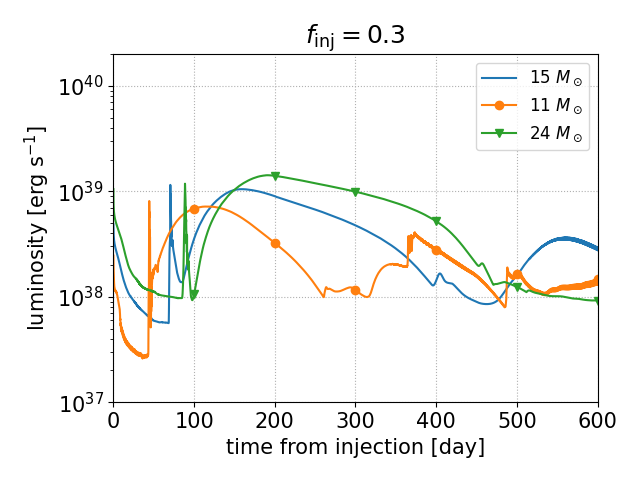}
    \end{minipage}
     \begin{minipage}[t]{0.5\hsize}
   \centering
    \includegraphics[width=\linewidth]{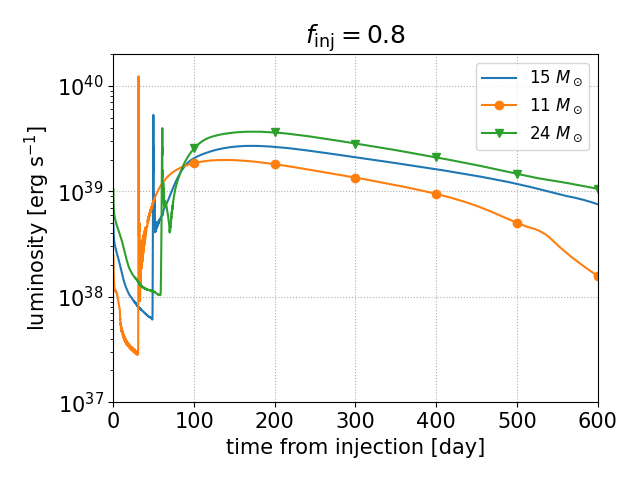}
    \end{minipage}
    \end{tabular}
 \caption{Bolometric light curves for models with different progenitors. The left panel show the case of $f_{\rm inj}=0.3$, and the right panel is the case for $f_{\rm inj}=0.8$.}
\label{fig:bolom_zams}
\end{figure*}

We model mass eruption with energy injection of $f_{\rm inj}=0.3$ and $0.8$, which correspond to the single-small and single-large models respectively in the main text. The light curves are shown in Figure \ref{fig:bolom_zams}. We note that for the $11\ M_\odot$ model with $f_{\rm inj}=0.3$, there is some instability seen in the light curve at late phases of $\gtrsim$ 400 days. This arises from the numerical instability caused by the large density discontinuity at the interface of the bound CSM and the star, after fallback of the bound CSM starts to dominate the emission. We thus focus the comparison on the shock breakout and the recombination-powered emission.

While the two components in the light curve of shock breakout and recombination emission are seen for all of the models, the light curves evolve generally faster and dimmer for a smaller ZAMS mass model. This is mainly found to be due to difference in the mass of the ejected envelope of around a factor 2 between the $11\ M_\odot$ and $24\ M_\odot$ models. The variation in both duration and peak luminosity of the recombination emission are both within a factor of 2. As these dependences appear to be degerate with the injected energy, the precursor light curves alone would be insufficient to infer both the progenitor and the injected energy. Independent information of the progenitor from archival data would be key to further understand the underlying energy injection processes. Images of massive stellar populations with sufficient resolution are available for sufficiently nearby galaxies within $20$--$30$ Mpc \citep{Smartt09}, which cover the typical distance of the supergiants we are targeting here.

\section{Dependence on Duration of Energy Injection}
\label{sec:duration}

\begin{figure*}
   \centering
    \begin{tabular}{cc}
     \begin{minipage}[t]{0.5\hsize}
    \centering
    \includegraphics[width=\linewidth]{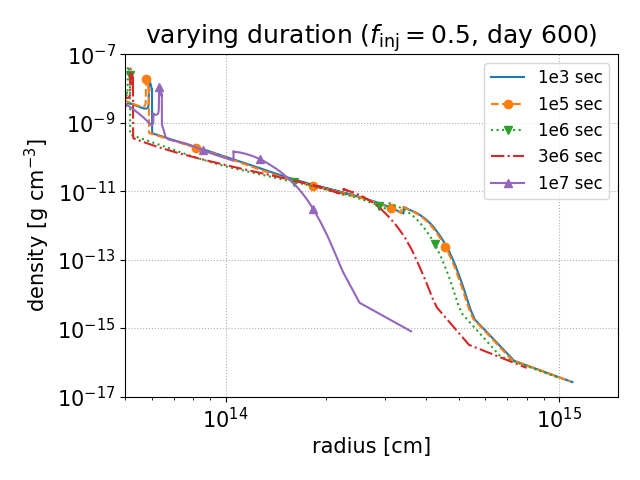}
    \end{minipage}
     \begin{minipage}[t]{0.5\hsize}
   \centering
    \includegraphics[width=\linewidth]{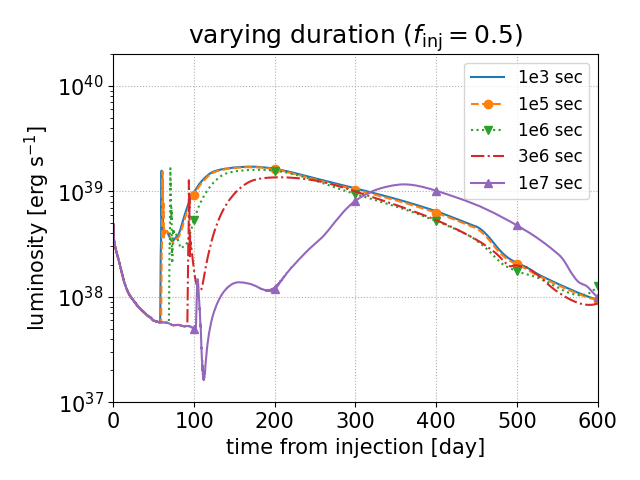}
    \end{minipage}
    \end{tabular}
 \caption{Results for changing the energy injection duration from the $10^3$ sec assumed in the main text. The left panel shows the density profile of the ejected material (CSM), and the right panel shows the light curves.}
\label{fig:tinj_comparison}
\end{figure*}

It is also important to check how our parameters would depend on the duration of the energy injection, which can be different for the various mechanisms proposed in the literature. Furthermore, \cite{Ko22} found that mass eruption can become much weaker when this duration is comparable to the dynamical timescale of the envelope. Here we study how the luminosity and the final CSM formation would depend on the duration of the injection.

We adopt the $15\ M_\odot$ progenitor as in the main text, and fix the total injected energy as $f_{\rm inj}=0.5$. We consider injection durations longer than the nearly instantaneous case adopted in the main text ($10^3$ sec), with values \{$10^{5}$, $10^6$, $3\times 10^6$, $10^7$\} sec. For all cases, injection is done with a constant luminosity, uniformly over the inner 10 cells.

The density profiles of the erupted material at the end of the simulations (600 days from energy injection) are shown in the left panel of Figure \ref{fig:tinj_comparison}. The density profiles for durations of $10^5, 10^6$ sec are nearly indistinguishable from that with the duration of $10^3$ sec. When the duration starts to become comparable to the dynamical timescale at the envelope $\sqrt{R_*^3/GM_*}\approx 8\times 10^6$ sec, the eruption starts to be affected by the gravity of the central star and the reduced energy supplied to the propagating shock. The resulting eruption is much weaker for the cases with durations of $3\times 10^6, 10^7$ sec. For these two cases, the unbound mass is drastically reduced to $0.14\ M_\odot$ and $5\times 10^{-4}\ M_\odot$ respectively, from the $0.35\ M_\odot$ for an instantaneous duration of $10^3$ sec. The drastic change of the unbound mass around this timescale qualitatively agrees with the results of \citet{Ko22}.

The corresponding light curves of the precursor emission are shown in the right panel of Figure \ref{fig:tinj_comparison}. Similar to the density profile, for injection durations of $\lesssim 10^{6}$ sec the precursor emission is almost insensitive to the duration. Significant difference is again seen for durations of $3\times 10^6$ and $10^7$ sec, where the weaker energy of the shock results in a delayed and weaker breakout pulse and a delayed peak of the recombination emission. Interestingly the peak luminosity is found to be similar for all the models, implying that the amount of envelope contributing to the recombination emission is insensitive to the injection duration.

The dependence of the mass loss on the energy injection duration was also studied in \cite{Leung21_fbot}, but for a helium star that lost all of its hydrogen envelope prior to energy injection. While they do not show the detailed light curve during the mass loss, in Figure 1 they obtain the luminosity profile at the end of the energy deposition phase (when they regard as core-collapse). For models that result in mass-loss the surface luminosity is inversely proportional to the injection duration. This hints that in contrast to our results with hydrogen-rich progenitors where hydrogen recombination powers the emission, for these stripped progenitors the precursor emission is powered by the energy injection itself and more directly reflects the injection history.

\end{appendix}

\bibliography{references}
\bibliographystyle{aasjournal}

\end{document}